\documentclass[conference,a4paper]{IEEEtran}
\usepackage{balance}
\usepackage{amsmath} 
\usepackage{graphicx}
\usepackage[nolist,nohyperlinks]{acronym}
\usepackage{multirow}
\usepackage{amsmath,amssymb,amsfonts}
\usepackage{algorithmic}
\usepackage[linesnumbered,ruled,vlined]{algorithm2e}
\SetKw{KWInitiate}{Initiate:}
\SetKw{KWCalculate}{Calculate:}
\SetKw{KWGenerate}{Generate:}
\SetKw{KWOptimize}{Optimize:}
\SetKw{KWCrossOver}{CrossOver:}
\SetKw{KWMutate}{Mutate:}
\usepackage[table,xcdraw]{xcolor}	
\usepackage{soul}
\usepackage{subcaption}

\SetCommentSty{mycommfont}

\SetKwInput{KwInput}{Input}                
\SetKwInput{KwOutput}{Output}              

\ifCLASSOPTIONcompsoc
 \usepackage[caption=false,font=normalsize,labelfont=sf,textfont=sf]{subfig}
\else
 \usepackage[caption=false,font=footnotesize]{subfig}
\fi
\begin{document}
\begin{acronym}
\acro{DRA}{Direct Radiating Array}
\acro{FoV}{Field of View}
\acro{SLL}{Side Lobe Level}
\acro{EIRP}{Effective Isotropic Radiated Power}
\acro{RF}{Radio Frequency}
\acro{GEO}{Geostationary Orbit}
\end{acronym}
%
\title{Genetic Algorithm-based Beamforming in Subarray Architectures for GEO Satellites}

\author{\IEEEauthorblockN{
Juan Andr\'es V\'asquez-Peralvo, Jorge Querol, Eva Lagunas, Flor Ortiz,\\ 
Luis Manuel Garcés-Socarrás, Jorge Luis Gonz\'alez-Rios, Victor Monzon Baeza, Symeon Chatzinotas.     
}                                     
\IEEEauthorblockA{
Interdisciplinary Centre for Security Reliability and Trust, University of Luxembourg, 1855 Luxembourg-Luxembourg\\ (e-mails: \{ juan.vasquez, jorge.querol, eva.lagunas, flor.ortiz, luis.garces, \\jorge.gonzalez, victor.monzon, symeon.chatzinotas\}@uni.lu).}

}



\maketitle

\begin{abstract}

The incorporation of subarrays in \ac{DRA} for satellite missions is fundamental in reducing the number of \ac{RF} chains, which correspondingly diminishes cost, power consumption, space, and mass. Despite the advantages, previous beamforming schemes incur significant losses during beam scanning, particularly when hybrid beamforming is not employed. Consequently, this paper introduces an algorithm capable of compensating for these losses by increasing the power, for this, the algorithm will activate radiating elements required to address a specific \ac{EIRP} for a beam pattern over Earth, projected from a \ac{GEO} satellite. In addition to the aforementioned compensation, other beam parameters have been addressed in the algorithm, such as beamwidth \ac{SLL}. To achieve these objectives, we propose employing the array thinning concept through the use of genetic algorithms, which enable beam shaping with the desired characteristics and power. The full array design considers an open-ended waveguide, configured to operate in circular polarization within the Ka-band frequency range of 17.7-20.2 GHz.\\

\end{abstract}

\vskip0.5\baselineskip

%

\section{Introduction}
The evolution of antennas in satellite communications exhibits a fascinating trend. Initially, the concept of employing a single feed per beam was adopted, necessitating one radiating element per beam and one satellite dish to cover specific areas over the Earth \cite{tomura2016trade}. Subsequently, the idea of utilizing the same reflector with different radiating elements to generate a coverage pattern was explored. The evolution progressed to the concept of multiple feeds per beam, enabling the generation of shaped beams using multiple radiating elements \cite{schneider2009multiple}. Following this, magnified arrays emerged, employing optics to amplify a radiating pattern using two reflector dishes and a small phased array antenna \cite{albertsen1987new}. The pinnacle of this evolution is the digital beamforming antennas called the ultimate antenna by A.J. Viterbi. This antenna is envisioned to be a DRA, which combines the concept of phased arrays with digital beamforming to control multiple beam characteristics, including control of beam shape, nulling, \ac{SLL}, and beam-steering.

Numerous methodologies have been investigated for beam synthesis using phased array antennas. These encompass analytical solutions employing Chebyshev, Taylor, Hamming, Hann, or other established taper distributions \cite{mailloux2018phased}, statistical methods \cite{schneider2009multiple}, \cite{lo1964mathematical}, deterministic techniques \cite{angeletti2009aperiodic}, non-deterministic techniques \cite{orchard1985optimising} \cite{vasquez2023flexible} or even machine learning models \cite{ortiz2023onboard}. The choice of a particular approach largely depends on the specific scenario under consideration. For example, array thinning lacks a deterministic or analytical solution that leads to a desired synthesized beam, rendering non-deterministic approaches like Genetic Algorithm (GA) particularly advantageous in such contexts.

This research introduces a non-deterministic approach to beam synthesis, which accounts for scanning losses within a subarray architecture in contrast to prior work. A \ac{GEO} serves as the focal point of this study, employing a \ac{DRA} operating in circular polarization within the frequency band of 17.7 - 20.2 GHz, requiring a beam projection over Earth with a diameter of 260 km at nadir. Additionally, this work addresses the design of the \ac{DRA} for this scenario and proposes a GA-based algorithm capable of addressing not only the aforementioned beam requisites but also other parameters like beamwidth in both azimuthal and elevation planes, alongside \ac{SLL} control, and \ac{EIRP}.

\section{Antenna Design}
\begin{figure}[ht!]
\centering
\includegraphics[width=0.8\linewidth]{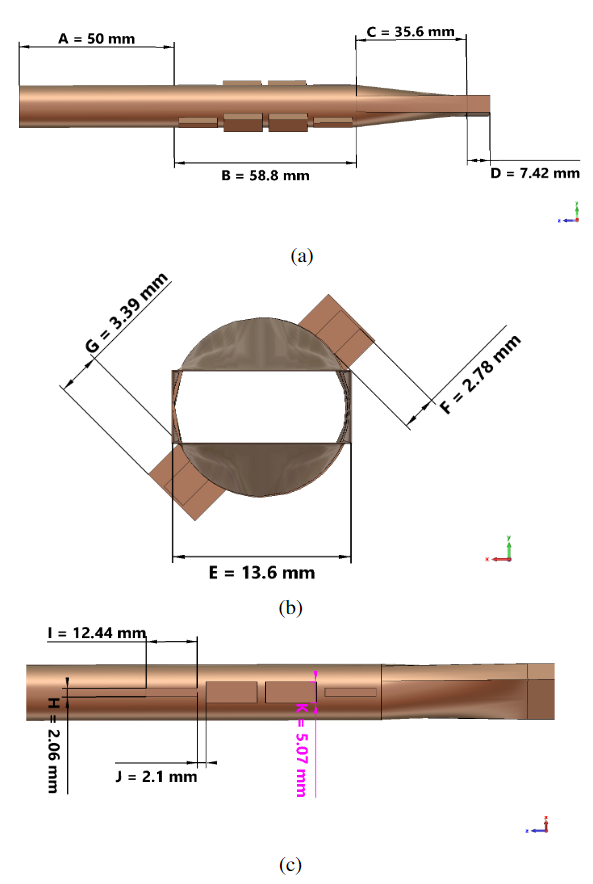}
\caption{Unit cell antenna layout and dimensions}
\label{fig:antenna}
\end{figure}

\begin{figure}[ht!]
\centering
\includegraphics[width=1\linewidth]{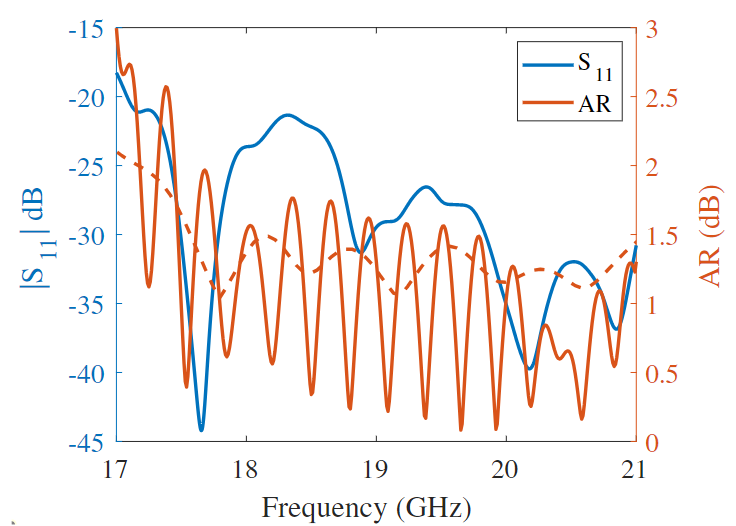}
\caption{S-Parameters results of the antenna depicted in Fig. \ref{fig:antenna}}
\label{fig:S_Paramters}
\end{figure}

\begin{figure}[ht!]
\centering
\includegraphics[width=0.9\linewidth]{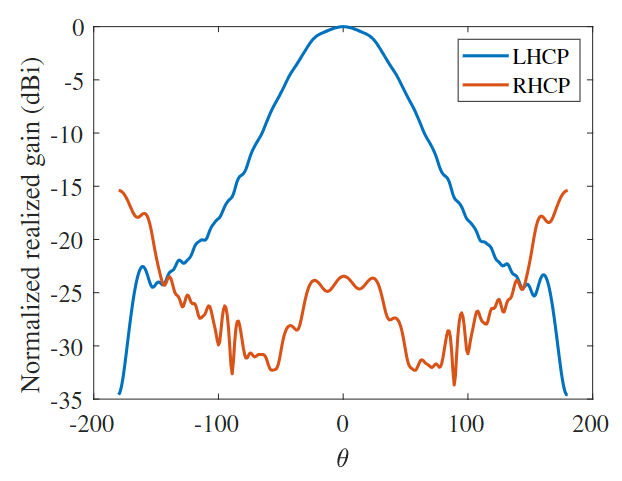}
\caption{LHCP and RHCP Radiation patterns of the antenna depicted in Fig. \ref{fig:antenna} }
\label{fig:RadPattern}
\end{figure}

 \ac{GEO}  missions face a significant disadvantage compared to other types of missions, primarily due to the substantial free space losses, which constrain the permissible losses that other components, such as the antenna, can incur. Given this limitation, the unit cell antenna designated for this scenario is an open-ended waveguide antenna, which exhibits significantly lower losses compared to alternative solutions like patch antennas or dielectric-based antennas. This antenna is composed of three segments: the open-ended waveguide itself, the groove polarizer to generate the circularly polarized waves, and the rectangular-to-circular transition that allows connecting the rest of the distribution network. Fig. \ref{fig:antenna} depicts the antenna's dimensions.
The simulation results, illustrated in Fig. \ref{fig:S_Paramters} and Fig. \ref{fig:RadPattern}, demonstrate that the antenna will radiate within the required frequency band, emitting Left-Hand Circular Polarized (LHCP) waves with a notably low cross-polarization component. This is evidenced by the $S_{11}$ parameter being lower than -10 dB, and an axial ratio less than 3 dB within the intended working frequency range.

\subsection{Array Dimensioning}
To determine the required antenna elements for this mission, we need to determine the coverage area required over the earth. Considering the Very High Throughput satellite configuration, where beams have to cover small areas to segment the total traffic, we have set the minimum beam diameter to 260 km or a coverage area $A_c$ = 53093 $\text{km}^2$  when the satellite is at nadir. It is worth mentioning that depending on the beam requirement to be addressed, the coverage area will be flexible. 

\begin{figure}[ht!]
\centering
\includegraphics[width=0.35\linewidth]{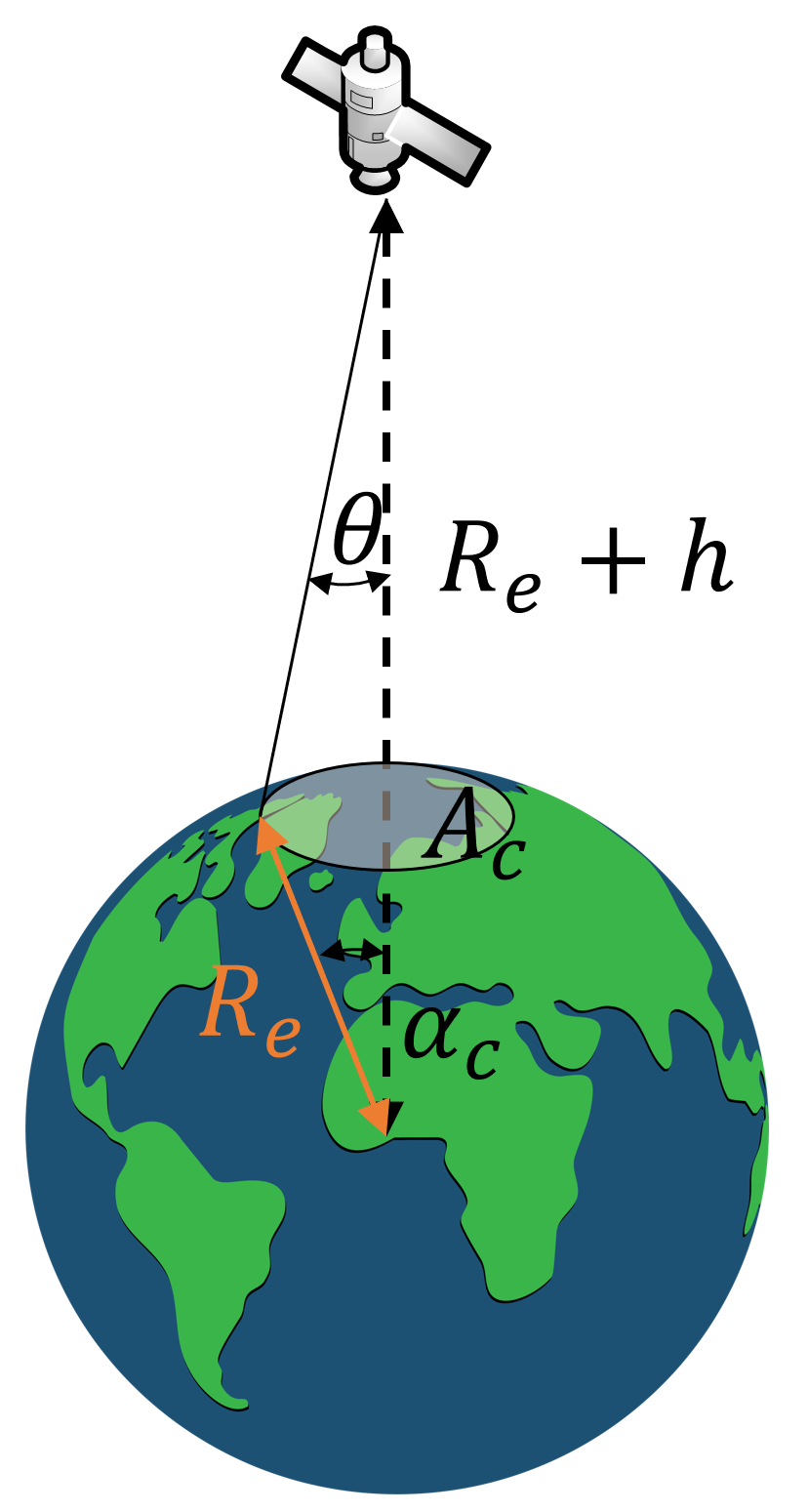}
\caption{Satellite scenario to determine the $\mathrm{\theta_{-3dB}}$viewed from the satellite antenna }
\label{fig:Scenario}
\end{figure}

The next step is to determine what is the 3 dB antenna beamwidth $\mathrm{\theta_{-3dB}}$ required to illuminate the required area; for this, we can use geometry to relate these two components, and the satellite location as illustrated in Fig. \ref{fig:Scenario}.

Utilizing $A_c$, we can estimate half of the angle of the satellite coverage $\alpha_c$, viewed from the earth center, by applying the spherical cap equation \eqref{Eq.SphericalCap}.
\begin{equation}
    \alpha_c = \cos^{-1}(-\frac{A_c}{2\pi R_e^2}+1)
    \label{Eq.SphericalCap}
\end{equation}
where $R_e$ is the radius of the earth and equal to 6317 km. Then, applying the law of sines, we can solve \eqref{Eq.Beamwidthangle} where $\theta$ is half of the angle corresponding to the $\theta_{\mathrm{-3dB}}$.
\begin{equation}
    \label{Eq.Beamwidthangle}
    \sin{\theta} = \frac{R_e}{R_e+h}\sin{(\theta+\alpha)}
\end{equation}
Using the satellite scenario data and applying previous equations, the antenna will need to provide $\theta_{\mathrm{-3dB}} = 0.41^\circ$.

The next step involves dimensioning the antenna number of elements per dimension $N_x$, $N_y$ using \eqref{Eq.NxNy},
\begin{equation}
    \label{Eq.NxNy}
    N_x = N_y = \frac{0.886 \lambda_0}{\eta \theta_{\mathrm{-3dB}}d}
\end{equation}
where $\lambda_0$ is the free space wavelength, and $\eta$ is the antenna efficiency, and $d$ is the antenna inter-element space.  Since the antenna will work at an operational frequency $f_0 =$19 GHz and assuming the maximum efficiency, the total number of elements $N_x = N_y$ = 144.

The previous results are impractical due to their demanding space, cost, and power requirements. This is mainly attributed to the necessity of equipping each antenna element with a full RF chain, comprising filters, amplifiers, ADC or DAC converters, and mixers. Consequently, employing subarrays emerges as a highly viable alternative.

The crucial factor in determining the number of subarray antenna elements lies in ensuring that no grating lobes intersect the antenna's \ac{FoV}) across the Earth's surface. To achieve this, it is essential to compute the FoV itself, as outlined in the equation referenced in Eq. \ref{Eq.FoV}, which is derived from the depiction provided in Fig. \ref{fig:Scenario}."
\begin{equation}
    \label{Eq.FoV}
    FoV = \tan^{-1}{\frac{R_e}{R_e+h}}
\end{equation}
Applying the previous formula gives a \ac{FoV} = $\pm 8.55 ^\circ$, which is the input for the inter-element space calculation in \eqref{Eq.inter-element}.
\begin{equation}
    \label{Eq.inter-element}
    d = \frac{\lambda_0}{2\sin{(FoV)}}
\end{equation}
As a result, the subarray must have antenna elements that do not exceed the size of $d$ = 3.5 $\lambda_0$ from each other to warranty scanning capabilities without having any grating lobes inside the \ac{FoV}. In this scenario, gathering 4$\times$4 elements will sufficiently address the previous requirements. Therefore, the total number of \ac{RF} chains is 36$\times$36, which highly reduces not just cost, mass, and power on the satellite but also computational complexity in the algorithm proposed in the next section.

To determine the the radiation pattern of the subarray and array, we can use the array factor formula depicted below:

\begin{align}
\label{eq:antenna_1}
    \mathrm{AF}=&\text{Ae}\sum_{m=1}^{M_x} {\rm e}^{j\left(m-1\right)\left(kd_x \sin \left(\theta \right)\cos \left(\phi \right)+{\beta_x} \right)}\times \nonumber\\  
      &\times\sum_{n=1}^{N_y} {\rm e}^{j\left(n-1\right)\left(kd_y \sin \left(\theta \right)\sin \left(\phi \right)+{\beta_y} \right)}\,,
\end{align}
where $M_x$ is the number of elements in the $x$-direction, $N_y$ is the number of elements in the $y$-direction, $k$ is the wave number, $d_x$ is the period in the $x$-direction, $d_y$ is the period in the $y$-direction, $\beta_x$ is the incremental phase shift in the $x$-direction, and $\beta_y$ is the incremental phase shift in the $y$-direction.
Then, the total field produced by the AF and the antenna element pattern, or if that is the case, the subarray pattern that we will call for both cases $E_{AE}$ is:
\begin{equation}
\label{eq:antenna_2}
E_{\rm T}=E_{\rm AE}\cdot{\rm AF}
\end{equation}

Finally, the total antenna gain is obtained as
\begin{equation}
    \label{eq:antenna_gain}
    G=e_{\rm eff}\cdot D(\theta,\phi)
\end{equation}
where $e_{\rm eff}$ is the antenna efficiency and the directivity $D(\theta,\phi)$ is defined as
\begin{align}
    \label{eq:antenna_4}
    D(\theta,\phi)=\frac{4\pi}{\int_{0}^{2\pi}\int_{0}^{\pi} E_{\rm T}^2\sin(\theta){\rm d}\theta {\rm d}\phi}\,.
\end{align}

\begin{figure}[ht!]
\centering
\includegraphics[width=1\linewidth]{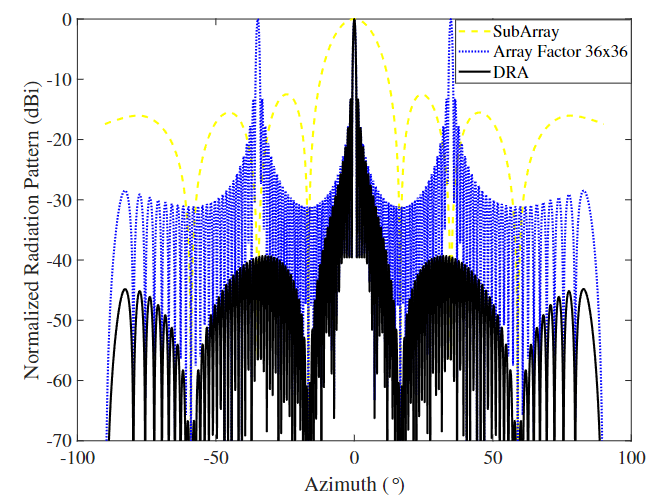}
\caption{Radiation pattern of the subarray, array factor of 36x36 elements separated 3$\lambda_0$, and the final radiation pattern of the DRA.}
\label{fig:RadaitionPatternA}
\end{figure}

To have a global view of the radiation pattern of each component, the patterns of the subarray, array factor of 36x36 elements separated 3$\lambda_0$, and the final radiation pattern of the \ac{DRA} are presented in Fig. \ref{fig:RadaitionPatternA}.

\section{Demand-Based Antenna Selection Algorithm}
The Genetic Algorithm stands out as the most suitable algorithm for discerning the active status of elements, and it has demonstrated remarkable efficiency in tackling optimization problems of this nature, as substantiated in \cite{orchard1985optimising}.
The algorithm aims to identify the appropriate sets of active elements required to generate a beam that satisfies the following constraints:
\begin{itemize}
\item Beamwidth in both Azimuth and Elevation to control to total capacity in certain area,
\item A minimum \ac{SLL} in the radiation pattern within the range of $\pm 8.5^\circ$ to avoid interference,
\item An specific \ac{EIRP} also to control the total capacity given in a certain region.
\end{itemize}

This concept extends the idea of array thinning by introducing additional constraints while accounting for power compensation due to the beam scanning process. Specifically, the scanning losses are exacerbated in this scenario at a certain pointing angle due to the subarray configuration. To mitigate these losses, the algorithm increases the number of active elements. In doing so, it still limits the search space with respect to other constraints, such as SLL and the required beamwidth.
In addition, to reduce the computational time and to avoid analyzing the \ac{SLL} in all directions, we have only calculated one-quarter of the weight matrix and replicated it to the rest of the quadrants to use symmetry in our favor.\\
The cost function created to address the desired beam will consider minimizing the error between the desired beamwidth, and the SLL in both planes and the required \ac{EIRP}. Cost functions formulation and the algorithm are depicted below:

\begin{equation}
\label{eq:CostFunction}
\min_{\vspace{1cm} W_{p\times q}^B}     \hspace{1cm}             Z_1(W_{p\times q}^B)+Z_2(W_{p\times q}^B)+Z_3(W_{p\times q}^B),
\end{equation}
where


\begin{equation*}
\label{eq:Eq2bF}
    \left\{
    \begin{aligned}
        & Z_1 =  \Bigg( \frac{|\mathrm{\theta_{-3dB_{Az_c}}^b}(W_{p\times q}^B)-\mathrm{\theta_{-3dB_{Az_o}}^b}|}{\mathrm{\theta_{-3dB_{Az_o}}^b}}+ \\ 
        & \hspace{1cm} \frac{|\mathrm{\theta_{-3dB_{El_c}}^b}(W_{p\times q}^B)-\mathrm{\theta_{-3dB_{El_o}}^b}|}{\mathrm{\theta_{-3dB_{El_o}}^b}} \Bigg)k_1\\  
        & Z_2 = \Bigg(\frac{|\mathrm{SLL_{Az_c}^b}(W_{p\times q}^B)-\mathrm{SLL_{Az_o}^b}|}{\mathrm{SLL{Az_o}^b}}+ \\
        & \hspace{1cm} \frac{|\mathrm{SLL_{El_c}^b}(W_{p\times q}^B)-\mathrm{SLL_{El_o}^b}|}{\mathrm{SLL{El_o}^b}}\Bigg)k_2\\
        &  Z_3 = \left(\frac{\mathrm{EIRP_c^b}(W_{p\times q}^B)-\mathrm{EIRP_o^b}}{\mathrm{EIRP_o^b}}\right)k_3, \\
    \end{aligned}
    \right.
\end{equation*}

\begin{algorithm}\DontPrintSemicolon
  
  \KwInput{$(\Lambda^B,\Phi^B)$, center of the beam in Latitude and longitude coordinates per beam,\\ \hspace{1cm}$\mathrm{\theta_{-3dB_{Az}^B}}$, beamwidth cut in azimuth per beam,\\ \hspace{1cm}$\mathrm{\theta_{-3dB_{El}^B}}$, beamwidth cut in elevation per beam,\\\hspace{1cm}$\mathrm{EIRP^B}$,  \ac{EIRP} per beam,\\\hspace{1cm}$\mathrm{SLL_{min}^B}$, SLL minimum per beam}
  \KwOutput{$W_{p\times q}^B$, Weight matrix based on previous inputs}
  \KwData{Set of possible configurations on Satellite considering system constraints}
   \KWInitiate{$\mathrm{Genetic}$ $\mathrm{Algorithm}$}\\
   \If{$\mathrm{counter} < \mathrm{counter_{\mathrm{max}}}$}
    {

  \KWCalculate{$W_{o}^B$}\\
  \KWCalculate{$\mathrm{radiation}$ $\mathrm{pattern}$, $\mathrm{\theta_{-3dB_{El}^B}}$, $\mathrm{\theta_{-3dB_{Az}^B}}$, $\mathrm{SLL_{min}^B}$, and $\mathrm{EIRP^B}$}\\
  \KWCalculate{$\mathrm{F}(Z_1 + Z_2 +Z_3 )$}\\
                  \eIf{$F<F_{\mathrm{min}}$}
                  {
                  $W_{p\times q}^B = W_{o}^B$\\
                  saves the optimal matrix\\
                  \textbf{break}
                  }
                  {
                 $\mathrm{counter} \gets \mathrm{counter} + 1$;\\
                  \KWCrossOver{$W_0^B$}\\
         \KWMutate{$W_0^B$}\\ 
                  
                  }

    }
\caption{Beam Forming Algorithm.}\label{algm:Beamforming}
\end{algorithm}

\section{Numerical Evaluation}
To demonstrate the output of the GA, we will consider the case of a satellite located at 13$^\circ$E, using a \ac{DRA} that has to address the specification described in Table \ref{table:table1}.
\begin{table}[ht!]
\centering
\begin{tabular}{|c|c|c|c|c|}
\hline
\rowcolor[HTML]{68CBD0} 
\textbf{Scenario} & \textbf{lat. lon ($^\circ$)} & \begin{tabular}[c]{@{}c@{}}$\mathbf{\theta_{-3dB}}$\\ ($\mathbf{^\circ})$\end{tabular} & \textbf{SLL (dB)} & \begin{tabular}[c]{@{}c@{}}\textbf{EIRP}\\ \textbf{(dBW)}\end{tabular}           \\ \hline
\rowcolor[HTML]{C0C0C0} 
1             & 39.3, -5.3                   & 0.9                      & \textgreater{}14  & 61.94                         \\ \hline
\rowcolor[HTML]{FFFFFF} 
2             & 49, 17.4                     & 0.9                      & \textgreater{}14  & 61.94                         \\ \hline
\rowcolor[HTML]{C0C0C0} 
3             & 39.3, -5.3                   & 1.4                      & \textgreater{}14  & \cellcolor[HTML]{C0C0C0}61.94 \\ \hline
\rowcolor[HTML]{FFFFFF} 
4             & 49, 17.4                     & 1.4                      & \textgreater{}14  & 61.94                         \\ \hline
\rowcolor[HTML]{C0C0C0} 
5             & 39.3, -5.3                   & 0.9                      & \textgreater{}14  & 49.93                         \\ \hline
\rowcolor[HTML]{FFFFFF} 
6             & 49, 17.4                     & 0.9                      & \textgreater{}14  & 49.93                         \\ \hline
\rowcolor[HTML]{C0C0C0} 
7             & 39.3, -5.3                   & 1.4                      & \textgreater{}14  & \cellcolor[HTML]{C0C0C0}49.93 \\ \hline
\rowcolor[HTML]{FFFFFF} 
8             & 49, 17.4                     & 1.4                      & \textgreater{}14  & 49.93                         \\ \hline
\end{tabular}
\caption{Simulations scenarios for the \ac{DRA} synthesis.}
\label{table:table1}
\end{table}

 The algorithm's output for these eight scenarios is presented in Table \ref{table:table2}. Analyzing the output table, we can infer the following. First, the SLL is over 14 dB in all cases, giving us confidence in the cost function. Second, as the  \ac{EIRP} decreases, the total number of active elements is also reduced, thus following the previously presented analysis. Third, in the case of the pointing 39.3$^\circ$, --5.3$^\circ$ the total number of active elements is equal or greater than for the other pointing angle. The previous goes in hand with the already described compensation the algorithm does when the scanning losses are high.
 \begin{table}[htp!]
\centering
\begin{tabular}{|c|c|c|c|c|c|}
\hline
\rowcolor[HTML]{68CBD0} 
\textbf{Scenario}             & \textbf{lat. lon ($^\circ$)}     & \begin{tabular}[c]{@{}c@{}}$\mathbf{\theta_{-3dB}}$\\ ($\mathbf{^\circ})$\end{tabular}      & \textbf{SLL (dB)}              & \begin{tabular}[c]{@{}c@{}}\textbf{EIRP}\\ \textbf{(dBW)}\end{tabular}            &\begin{tabular}[c]{@{}c@{}}\textbf{Active}\\ \textbf{Element}\end{tabular} \\ \hline
\rowcolor[HTML]{C0C0C0} 
1                         & 39.3, -5.3                       & 0.897                         & 17.319                         & 62.204                         & 484                                                                   \\ \hline
\cellcolor[HTML]{FFFFFF}2 & \cellcolor[HTML]{FFFFFF}49, 17.4 & \cellcolor[HTML]{FFFFFF}0.899 & \cellcolor[HTML]{FFFFFF}17.545 & \cellcolor[HTML]{FFFFFF}61.912 & 468                                                                   \\ \hline
\rowcolor[HTML]{C0C0C0} 
3                         & 39.3, -5.3                       & 1.331                         & 15.824                         & \cellcolor[HTML]{C0C0C0}58.053 & 300                                                                   \\ \hline
\cellcolor[HTML]{FFFFFF}4 & \cellcolor[HTML]{FFFFFF}49, 17.4 & \cellcolor[HTML]{FFFFFF}1.345 & \cellcolor[HTML]{FFFFFF}16.642 & \cellcolor[HTML]{FFFFFF}55.968 & 236                                                                   \\ \hline
\rowcolor[HTML]{C0C0C0} 
5                         & 39.3, -5.3                       & 0.918                         & 15.319                         & 52.369                         & 156                                                                   \\ \hline
\cellcolor[HTML]{FFFFFF}6 & \cellcolor[HTML]{FFFFFF}49, 17.4 & \cellcolor[HTML]{FFFFFF}0.900 & \cellcolor[HTML]{FFFFFF}14.217 & \cellcolor[HTML]{FFFFFF}52.369 & 156                                                                   \\ \hline
\rowcolor[HTML]{C0C0C0} 
7                         & 39.3, -5.3                       & 1.388                         & 16.129                         & \cellcolor[HTML]{C0C0C0}50.379 & 124                                                                   \\ \hline
\cellcolor[HTML]{FFFFFF}8 & \cellcolor[HTML]{FFFFFF}49, 17.4 & \cellcolor[HTML]{FFFFFF}1.369 & \cellcolor[HTML]{FFFFFF}15.101 & \cellcolor[HTML]{FFFFFF}49.495 & 112                                                                   \\ \hline
\end{tabular}
\caption{Simulations results of the antenna synthesis.}
\label{table:table2}
\end{table}

In the case of the pointing error, we can see that it is non-existent; this is expected since the algorithm is not optimizing the beam pointing because the calculation is done using the progressive phase shift formula. Conversely, we have errors in beamwidth and \ac{EIRP} that need to be further analyzed. The error in beamwidth reaches its peak for the third beam generation, being around 4.95\%. Some degree of errors are expected for the beamwidth calculation. In the case of the \ac{EIRP}, the errors are more important and located in beams 3-6. Doing a further analysis of these beams, we can see that the requirements were a combination of different pointing angles with high \ac{EIRP} with high beamwidth and low \ac{EIRP} with low beamwidth. The algorithm finds previous combinations difficult to achieve due to the inverse relation between\ac{EIRP} and beamwidth; therefore, the algorithm will do its best for some combinations, but there will be some errors in the extracted beam parameters.

\begin{figure}[ht!]
\centering
\includegraphics[width=0.4\linewidth]{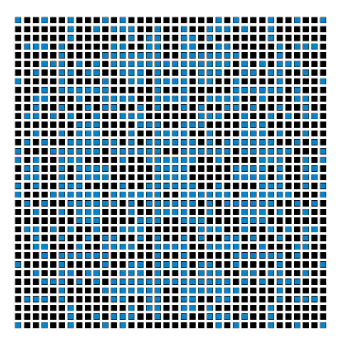}
\caption{Active and not active antenna elements obtained after the optimization to address the requirements of the desired scenario 1. Blue = activated, Black = deactivated.}
\label{fig:thinnedArray}
\end{figure}

\begin{figure}[htp!]
     \centering
        \begin{subfigure}[b]{\linewidth}
         \centering
         \includegraphics[width=\textwidth]{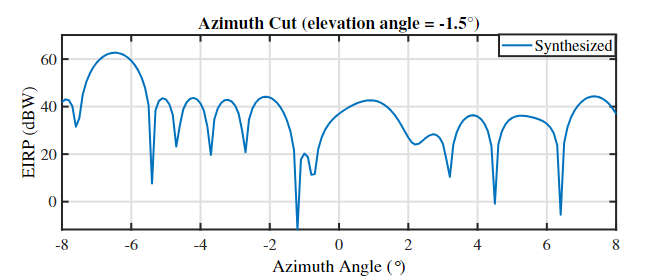}
         \caption{}
         \label{fig:Beam1a}
     \end{subfigure}
     \hfil
     \begin{subfigure}[b]{\linewidth}
         \centering
         \includegraphics[width=\textwidth]{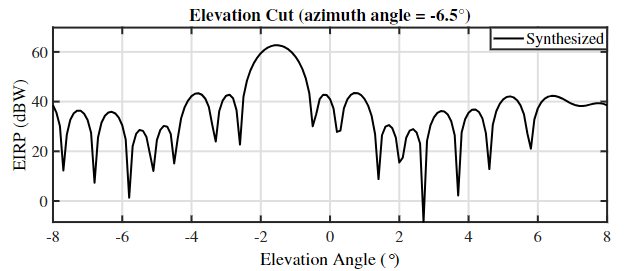}
         \caption{}
         \label{fig:FigBeam1b}
     \end{subfigure}
      \hfil
     \begin{subfigure}[b]{\linewidth}
         \centering
         \includegraphics[trim=0 0 0 0,clip,width=\textwidth]{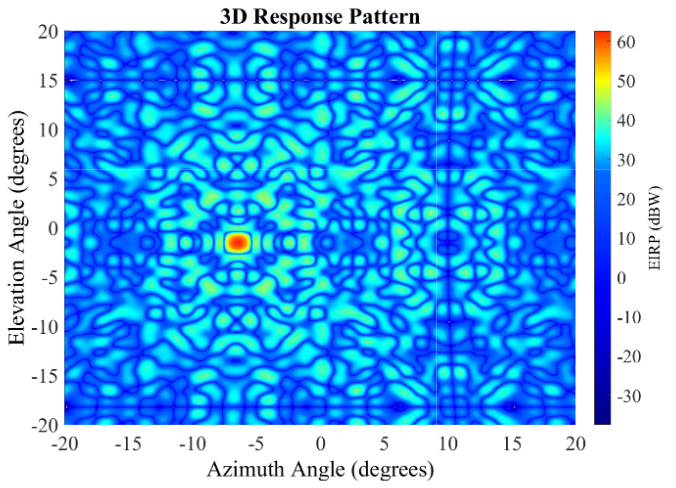}
         \caption{}
         \label{fig:FigBeam1c}
     \end{subfigure}
    \caption{Synthesized radiation pattern cuts to address the beam requirements described in Table \ref{table:table2} Scenario 1. a) Azimuth cut. b) Elevation cut. c) 3D Radiation pattern \ac{EIRP}} 
    \label{fig:Beam1}
\end{figure}

We will take the outcome for scenario 1 to disseminate the algorithm's results. The simulation time it takes to generate one beam is around 584.2 seconds in an Xeon core-based laptop. The synthesized \ac{DRA} is visibly depicted in Fig. \ref{fig:thinnedArray}, and the cuts and 3D beam are illustrated in Fig. \ref{fig:Beam1}.

An interesting aspect of the synthesized beam is that the pointing angle is obtained by the position of the satellite over the service area using a coordinate transformation to transform it in terms of azimuth and elevation. 

\section{Conclusion}
A beam synthesis that uses a Genetic Algorithm has been devised to generate an optimized weight matrix, with inputs including beamwidth in both planes, overall \ac{SLL} s, and effective isotropic radiated power for a \ac{DRA} tailored for satellite communications. This design framework, detailed through dimensions, design specifications, and simulation results, adopts a \ac{GEO} satellite as a case study. The \ac{DRA} architecture employs open-ended waveguides paired with a groove polarizer to attain the requisite circular polarization. Notably, the proposed \ac{DRA} harnesses sub-arrays to diminish the count of \ac{RF} chains. This strategy alleviates computational demands and simulation duration and, in practice, paves the way for a cost-efficient, lightweight, and energy-saving practical implementation. Moreover, this algorithm's versatility could be applied in multi-beam scenarios.


\section*{Acknowledgment}
This work has been supported by the European Space Agency (ESA) funded under Contract No. 4000134522/21/NL/FGL named "Satellite Signal Processing Techniques using a Commercial Off-The-Shelf AI Chipset (SPAICE)". Please note that the views of the authors of this paper do not necessarily reflect the views of ESA.



%
\bibliographystyle{IEEEtran}
\bibliography{References.bib}

\end{document}